# Towards Identifying the Systems-Level Primitives of Cortex by In-Circuit Testing

Leslie G. Valiant

*valiant@seas.harvard.edu*

Harvard University

July 22, 2018

**Summary**

The hypothesis considered here is that cognition is based on a small set of systems-level computational primitives that are defined at a level higher than single neurons. It is pointed out that for one such set of primitives, whose quantitative effectiveness has been demonstrated by analysis and computer simulation, emerging technologies for stimulation and recording are making it possible to test directly whether cortex is capable of performing them.

1. **Introduction**

In both biology and computer science the virtue of modularity in complex systems is widely understood. In contrast, the idea that complex systems need to have many levels of description, while well appreciated in computer science, has been much less discussed in biology. We believe that for a computational theory of cortex to be useful, it would need to give an account of algorithmic tasks at levels intermediate between those of neurons and behavior. Just as we cannot understand the workings of a cell phone by studying its operations only at the single bit level, we cannot expect to understand the brain at the single neuron level.

The suggestion we explore here is that cognitive computation in cortex is built on a small collection of base primitives at the systems level. This contrasts with Hebbian plasticity (Hebb, 1949), which is a single primitive at the single neuron level. Here we discuss one candidate for such a collection of base primitives, one distinguished by the existence of demonstrations, by analysis and computer simulations, that it offers a scheme in which the capacity of cortex for basic cognitive tasks can be given a quantitative explanation (Valiant 1994, 2005, 2006; Feldman & Valiant 2009). This collection consists of: association, memorization, inductive learning, and hierarchical memory assignment, all suitably defined. The purpose of this note is to point out that because of technological advances over the last decades, especially in optogenetics (Miyawaki *et al*. 1997, Boyden *et al*. 2005, Deisseroth *et al*. 2006, Shemesh *et al*. 2017), it is now becoming feasible to test systematically whether cortex can indeed perform these primitives.

These four primitives are defined below in terms of their actions on *randsets*, random sets of neurons of a certain size. The nature of the proposed experiments is analogous to "in-circuit testing" of a traditional electronic printed circuit board, where instead of testing the input-output behavior of the whole board, the testing is done on smaller parts, using only probes. In the proposed experiments there will be no sensory stimuli or behavioral responses – the experiments will consist entirely of stimulating and recording from appropriate sets of neurons. The goal is to find out what changes in computational function in a subcircuit of cortex can be achieved by specific neural stimulation. Existing experimental evidence that the firing patterns of cortical neurons can be altered by suitable behavioral training (Jackson *et al.* 2006, Rebesco *et al.* 2010) is relevant and encouraging, but does not directly address the question of which computational primitives are being executed. Behaviors prompted by sensory inputs cause all kinds of activity in the brain extraneous to the execution of any one primitive. The in-circuit methodology is suggested to isolate the operation of a subcircuit as much as possible from such interference.

The goal therefore is to test, by direct neural stimulation and recording, whether cortex is capable of performing certain basic computational tasks. The paradigmatic first such task is "association of a set $A$ to set $B$", abbreviated as $A \rightarrow B$, which here is defined as follows: for neuron sets $A$ and $B$ we want to induce the new functionality that, after the training, whenever set $A$ is stimulated then the newly modified circuit will cause set $B$ to become active also. Training here would involve stimulating $A$ and $B$ according to some timed protocol. Testing would involve stimulating $A$ again and now recording from the set $B$ that had been stimulated in training. This particular task may be viewed as a direct systems level analog of Hebbian plasticity.

We note that our notion of randsets is not the same as the notion of assemblies, in the sense of Hebb (1949). That traditional notion of assemblies is associated with the idea that the neurons in these assemblies are better connected to each other than to other neurons (e.g. Carillo-Read et al. 2017). Randsets account for the claimed computational capabilities simply by virtue of being randomly chosen, and probably with weak criteria of randomness being sufficient. There is *no* requirement that they be well connected to each other. The theory basically asserts that for random sets in a randomly connected network there will be a large enough bandwidth of reliable communication from one randset to another, that simple local updates at the neurons that result from training will be sufficient to realize the claimed functionalities. There is no comparable theory that shows that the high interconnectivity assumption of traditional assemblies provides any comparable computational capabilities. Fortunately, also, it is the randset theory that appears to be one that is readily testable experimentally for the sets of operations we contemplate, since random sets are more easily identified than highly interconnected ones. (Of course, it may be that our randsets, once selected, do become more strongly connected to each other over time, but that is not our concern here.)

An experiment relevant to what we are contemplating here is reported in Seeman *et al.* (2017), where the setting up of such associations was attempted by electronic stimulation, but did not succeed reliably. The experiment was done for one value of *r*. Possible interpretations are that the value of *r* used there was too small, or that it was not a part of cortex that performs this function. We are suggesting a systematic exploration of the relevant parameter space, and in particular the set size *r*. Another relevant result, one obtained by optogenetic means, is by Carrillo-Reid *et al*. (2016), which can be viewed as also an example of "in-circuit" testing, but the task considered there was pattern completion, as in auto-associative nets (e.g. Hopfield, 1982). We believe that pattern completion is a less fundamental operation for our purposes than the ones we consider here, since there is no clear definition of what it should be, and any reasonable system offers some version of it as an epiphenomenon.

2. **The Four Tasks and their Potential Validation**

Each instance of the tasks described here can be induced by an associated training protocol, and is realized as local changes in the neurons within the given randomly connected network, giving rise to a subcircuit having the functionality of that task instance. Computer simulations (Feldman & Valiant, 2009) show that thousands of instances of such tasks can be realized with realistic network parameters without having the continued effectiveness of the earlier ones acquired degraded by later ones. We note that having multiple task types, which between them may be adequate to form a basis for cognition, is a much more exacting requirement than single-task models such as traditional "auto-associative nets" which perform the single task of pattern completion. The longer term motivation is that of identifying a "cognitively adequate'' set of primitives, perhaps our four forming the core, and showing that higher level cognitive operations can be efficiently implemented in terms of these.

The following detail the four tasks that we are suggesting as the core, and for which we are here suggesting validation by in-circuit testing. An example of some more detailed specifications still, for one incarnation, is described by Feldman and Valiant (2009).

**Association:** For sets *A*, *B* train so that result is: If in future *A* is stimulated then *B* (but not random other neurons) will become active.

**Supervised Memorization of Conjunctions:** For sets *A*, *B*, *C* train so that the result is: If in future *A* and *B* are stimulated then *C* will become active also (but not if just one of *A*, *B* is stimulated.)

**Inductive Learning of Simple Threshold Functions**: For sets *A*, *B*, *C*, *D* (say) train so that the resulting circuit generalizes a classification at target *A*, beyond the examples seen, according to a linear separator on the variables *B, C, D*. In particular, if in future some subset of *B*, *C*, *D* is stimulated then the target *A* will become active according to whether some linear separator consistent with the examples holds. For example such a linear separator may be $B + C + 2D \geq 2$, where now these variables have a 1/0 interpretation corresponding to whether the randsets are active or not.

**Hierarchical Memory Formation:** For sets *A*, *B*, train so that result is: a set *C* (not specified by the experimenter) has been allocated by the computation internally so that the effect "supervised memorization of conjunctions" is achieved for *A*, *B* and that, now unknown, *C*. This realizes the fundamental psychological task of "chunking", where a new compound concept, here *A & B*, becomes equal citizen in the circuit with earlier ones.

While the first three of these tasks can be tested directly by in-circuit testing, the last task appears more difficult. One possible approach is to glean information using the timing mechanism at work.

Verifying *how* these tasks are realized in cortex, if indeed they are, is a further area of challenge. The suggestion has been made (Valiant, 2012) that the hippocampal system is responsible for determining the identity of the randset *C* in hierarchical memory formation, for given *A* and *B*. Such a statement, which assigns responsibility for performing a particular task to a particular brain area, may be easier to test.

3. **General Considerations**

**How large sets?** In the analysis the parameter *r* is dependent on the number of neurons in the system, the number of connections, the synaptic strengths, and the algorithms used for realizing the tasks. For the association task, for example, having *r* too small will fail to train the target *B* as desired, while having *r* too large will train too many spurious neurons in addition. Hence, we expect that considerable experimentation may be needed to find the right value of *r* in any one cortical area. Note that in this case of associations, the size of the source set *A* appears to be critical, but this is the easier set in that it only needs stimulation. The size of the target set *B*, which has to be both stimulated and recorded from, may be much less critical.

**Which brain areas?** To test our theory one needs to perform the experiment in a cortical area where the corresponding task is implemented in biology. We have behaving mammals in mind. It may be that primates are the best at some of these tasks. There is some uncertainty as to which parts of cortex perform these various tasks, and some experimentation there may be necessary.

**How sets chosen?** Since the hypothesis requires that the experiment work for randomly chosen sets of a certain size in the appropriate brain areas, it is reasonable to go with whatever bias the experimental technique used imposes. Neurons in particular layers may need to be found. The theory aims to explain how large numbers of task instances can be processed. Hence, experiments need to be done in parts of cortex that are large enough that it is plausible that so many tasks are indeed supported. The various sets *A*, *B*, *C*, etc., may be in the same cortical area, or in different areas.

**What timing protocols for training?** Timing may be critical. When stimulating a pair of sets *A*, *B* for association we need an asymmetric protocol, such as *B* being stimulated with one millisecond delay after *A*.

**Comparisons?** Comparative studies for different cortical areas and different species, including humans, would be of great interest.

## 4. Conclusion

The hypothesis to be tested is that cortex is able to perform a specific set of tasks on random sets of neurons. The new opportunity is offered by the remarkable recent advances in stimulation and recording technologies. The experiments would determine *whether* these tasks can be performed in appropriate cortices, but not how. However, positive results would demonstrate that the brain has impressive computational capabilities that are relevant to cognitive computation. These systems-level capabilities, if present, are unlikely to have arisen by accident. It would be a remarkable coincidence if the brain had these capabilities but did not use them.

## 5. Acknowledgement

This work was supported in part by a grant from the National Science Foundation CCF 1509178.